# Peripheral Nervous System Responses to Food Stimuli:

## *Analysis Using Data Science Approaches*

Moranges M[1,2], Plantevit M[2], Bensafi M[1]

[1] Lyon Neuroscience Research Center, CNRS, INSERM, Université Lyon1

[2] Laboratoire d'Informatique en Image et Systèmes d'information, Université Lyon1, CNRS, LIRIS UMR 5205, F-69622, Lyon, France

**Abstract**

In the field of food, as in other fields, the measurement of emotional responses to food and their sensory properties is a major challenge. In the present protocol, we propose a step-by-step procedure that allows a physiological description of odors, aromas, and their hedonic properties. The method rooted in subgroup discovery belongs to the field of data science and especially data mining. It is still little used in the field of food and is based on a descriptive modeling of emotions on the basis of human physiological responses.

# 1. Introduction

The measurement of emotions is a real challenge for basic and applied research, especially in the field of food. Often, the analysis of emotions in relation to food stimuli is performed in a multimodal manner by combining subjective measures (valence or self-reported hedonic preferences) and more objective measures such as peripheral nervous system activity. Here, the challenge is to measure hedonic preferences related to odors, taste, visual aspects, texture of food products, or of the food as a whole, and to associate these verbal and declarative responses with more objective measures of emotion such as electrodermal or cardiac activity *(1–5)*.

To date, most studies that have attempted to understand the physiological underpinnings of hedonic preferences to food sensory stimuli have used standard statistical approaches, such as comparing psychophysiological responses across different conditions (e.g., pleasant or unpleasant). Primarily used statistical methods included nonparametric tests (e.g., Wilcoxon, Kruskal Wallis) *(6, 7)* and parametric tests (analysis of variance, ANOVA) *(8–11)* depending on study design, data normality, and/or sample size.

Today, complementary approaches from data science can shed different light on these data by allowing researchers in the field to test descriptive and predictive models of the relationship between subjective preferences and psychophysiological responses. These approaches may be particularly well-suited to the format of data generated in emotion and food science as they take into account large, heterogeneous, and complex data. Indeed, data science can be applied to different types of data and signals (heart rate, skin conductance, chemical data, MRI, etc.). It allows the treatment of a large number of stimuli (thousands/billion) and a large number of attributes in parallel (in our example: four dimensions of the skin conductance will be analyzed together).

Data science is a general term used to describe the various aspects of data processing, with the aim of extracting meaningful information and relevant knowledge *(12)*. It includes data preprocessing (cleaning, normalization, discretization, etc.), data modeling, and data visualization. Within data modeling, we distinguish two main families of algorithms: those referring to machine learning, which are often predictive, and those belonging to data mining, which are often descriptive. Thus, whereas machine learning allows us to predict a variable (e.g. hedonic preference) on the basis of one or more variables (e.g. physiological responses) (see Note 1), data mining allows us to build descriptive models. The latter are often explanatory in the sense that they explain by explicit association rules how a pleasant food flavor is characterized physiologically compared to another less pleasant food for example.

The present protocol, inspired from a previous study *(13)*, aims to provide researchers and students in the field of food science with a framework for using these data mining methods, especially subgroup discovery methods, in the context of research on food-related emotions. We will use electrodermal responses (e.g., skin conductance or SC) and ratings of pleasantness collected from human individuals in response to various olfactory stimuli as an example of data *(14)*. Using these example data, we provide the user with a step-by-step protocol that allows for a physiological description of odors and aromas characterized by their hedonic value.

## 2. Materials

1. A computer (*Specifications: Windows 7 or later, Mac OS X 10.11 or higher, or Linux RHEL 6/7*)
2. Python 3 SDK software: It is an interpreted, multi-paradigm, and multi-platform programming language (see Note 2). It supports structured, functional and object-oriented imperative programming.
3. A Python development environment: Jupyter notebook; to be able to view the notebook created and read it as a tutorial.
4. The following libraries should be installed on Python:
   a. sickit-Learn (version = 0.24.1) *(15):* it is a library that offers various possibilities in terms data science methods
   b. matplotlib (version = 3.1.1): to handle graphical representation,
   c. pandas (version = 1.2.3): to visualize and to manipulate tabular data,
   d. pysubgroup (version = 0.7.2) *(16)*: offers several datamining algorithms.

   One can install Jupyter notebook and all of the above libraries by installing a Python distribution named Anaconda and by installing the pysubgroup package with the following command : *pip install pysubgroup.*
5. A database with subjective and psychophysiological data. As an example, we will use a dataset that combines SC responses to pleasant and unpleasant odors (see Licon et al. *(14)*). The dataset is downloadable from a public repository (Dataset name: "*PsychophysioDataset.xlsx"*, available at https://github.com/mmaelle/Psychophysio-Analysis). Each dataset row is an observation for a specific odor with its subjective pleasantness (rated using a scale from 1, very unpleasant, to 9, very pleasant) and its associated SC response described by four parameters (amplitude, latency, rise time, and number of events in a period following stimulus onset). The database contains 2398 observations (109 observations for 22 individuals).

# 3. Methods

The protocol contains two steps developed below: (1) Preprocessing, and (2) Data mining analysis. An overview of the whole analytical process is illustrated in Fig. 1.

*1. Preprocessing*

Preprocessing involves all the operations that precede the analysis step. Preprocessing is an important step that should not be neglected: it prepares the dataset so that it is as clean (without error), as simple as possible, and is adapted to the algorithms that will be used in the following analysis (modeling and data format compatible with the input parameters of the algorithm). To do this, noisy, outlier or irrelevant data must be corrected or rejected so as not to bias the study.

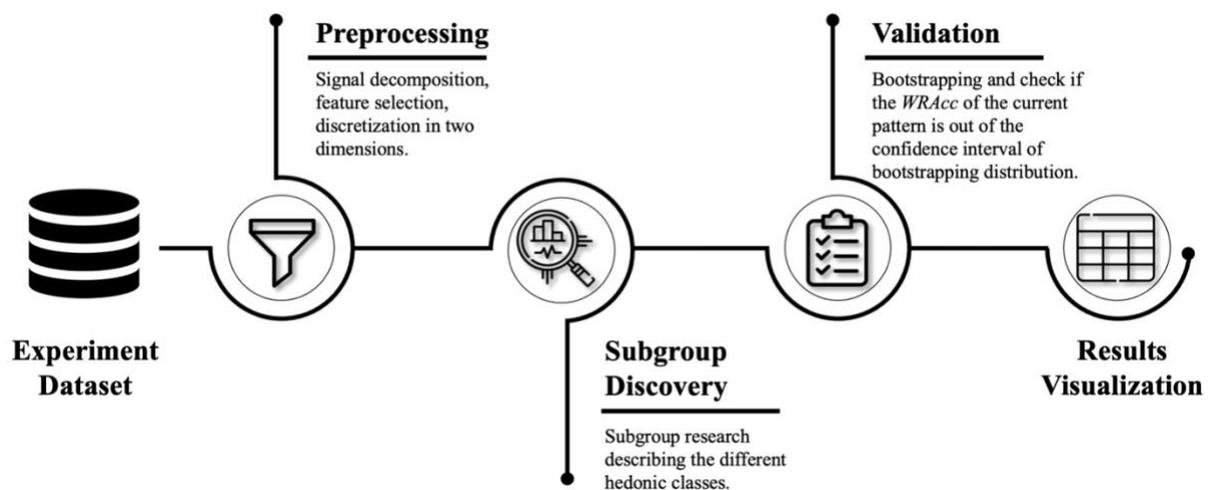

*Fig. 1. Steps of the workflow.*

## 1.1. Skin Conductance Decomposition and Data Importation

In a first step, we start by importing the data — available in the excel file "PsychophysioDataset.xlsx" — into a python dataframe. The data set contains the following information: participant number, CID (Compound Identification number of the odorant), latency, rise time, amplitude, number of events in a time window following stimulus onset, and ratings of odor pleasantness. Note that if no peak has been identified, the attributes latency, amplitude, and rise time are equal to 0. Moreover, the number of events is calculated by subtracting the number of events before the presentation of the odor from the number of events after the presentation of the odor (the value can therefore be negative). For more details, refer to the second session in Licon et al (14).

```
import pandas as pd

data = pd.read_excel('PsychophysioDataset.xlsx',
            sheet_name='Psychophysio')

print(data.head())
```

The preceding command displays the first lines of the Psychophysio dataset (Table 1).

|   | Subject | Stimulus | Latency | Rise-Time | Amplitude | Events | Pleasantness |
|---|---------|----------|---------|-----------|-----------|--------|--------------|
| 0 | 1 | 1 | 2.98046875 | 3.01171875 | 0.19736884 | 1 | 5 |
| 1 | 1 | 2 | 1.88671875 | 6.46875 | 0.91188065 | 2 | 4 |
| 2 | 1 | 3 | 2.0859375 | 4.74609375 | 0.27328656 | 1 | 0 |
| 3 | 1 | 4 | 1.69921875 | 4.515625 | 0.61869071 | -2 | 5 |
| 4 | 1 | 5 | 0.80078125 | 3.4375 | 0.08303131 | 0 | 3 |

**Table 1.** Initial dataset

1.2. Feature Selection

In a second step, irrelevant data (such as CID) that are not directly helpful to answer our question can be removed as follows.

```
df = data[['Subject', 'Latency', 'Rise-Time', 'Amplitude','Events',
'Pleasantness']].copy()
```

Incomplete data such as trials for which participants did not provide perceptual scores or for which SC data are missing can also be rejected.

```
df = df.drop(list(df[df['Pleasantness']==0].index))
nan_cols = [i for i in df.columns if df[i].isnull().any()]
for c in nan_cols :
    df = df.drop(list(df[df[c].isnull()].index))
```

Based on your data set, you will decide which variables are of interest and which should not be included in or should be rejected from your analysis.

1.3. Discretization

In a third step, we seek to label our data as "pleasant odors" and "unpleasant odors" for pleasantness. We will therefore discretize the self-reported scores into two classes. Having two discrete classes instead of scores allows one to deal with binary responses and to clearly separate pleasant from unpleasant odors on the one hand and weak from strong odors on the other hand. We choose here to discretize the scores using the clustering algorithm called *KMeans (17)* because this algorithm allows one to get rid of the subjectivity linked to the different scoring strategies (See Note 3). To obtain 2 groups (unpleasant and pleasant), we use *K=2*.

We thus obtain the preprocessed dataset presented in Table 2. It is available in csv format in the file *"PsychophysioPreprocessed.csv"*.

|   | Subject | Latency | Rise-Time | Amplitude | Events | Pleasantness_class |
|---|---|---|---|---|---|---|
| 0 | 1 | 2.98046875 | 3.01171875 | 0.19736884 | 1 | pleasant |
| 1 | 1 | 1.88671875 | 6.46875 | 0.91188065 | 2 | unpleasant |
| 2 | 1 | 2.0859375 | 4.74609375 | 0.27328656 | 1 | pleasant |
| 3 | 1 | 1.69921875 | 4.515625 | 0.61869071 | -2 | unpleasant |
| 4 | 1 | 0.80078125 | 3.4375 | 0.08303131 | 0 | Pleasant |

**Table 2.** Preprocessed dataset

## *2. Data Mining Analysis (Subgroup Discovery Analysis)*

2.1. General Information

We propose to use a data mining approach based on a subgroup discovery (SD) algorithm *(18)*. SD allows the discovery of patterns that are discriminating for a target class. Indeed, it finds population subgroups that are statistically "most interesting" from a population of individuals (or items). These subgroups are identified by conditions on the descriptive attributes. In this regard, we seek to obtain the largest possible subpopulations that have the most unusual statistical distributional characteristics.

For example, we obtain a pattern described by the rule "*Event==0.0 AND Rise-Time>=5.35 → unpleasant*". The conditions on the physiological attributes form the property of interest "*Event==0.0 AND Rise-Time>=5.35*". This property describes the subgroup identified as having exceptional behavior: the distribution of the target (unpleasant) is high in this subgroup compared to the rest of the dataset. This means that having a rise time greater than or equal to 5.35 seconds and at the same time a constant number of skin conductance peaks is significantly more present for the "unpleasant" trials than for the other so-called "pleasant" trials. This pattern is illustrated by the dataset depicted in Fig. 2. The subgroup consists of the three colored rows in the dataset that verify the conditions present in the rule. In the dataset, rows belonging to the target class form the "positive dataset" (in bold in Fig. 2) and the other rows form the "negative dataset" (in italics in Fig. 2). Note that in the "positive dataset", the items that are part of the subgroup are called "positive subgroup" (or "positives_sg").

**Fig. 2.** *Dataset example for the pattern described by the "Event==0.0 AND Rise-Time>=5.35 → unpleasant" rule.*

2.2. Processing

First, we import the data set, "pandas" library for viewing and manipulating tabular data and a library for subgroup discovery analysis called "pysubgroup".

```
import pandas as pd
import pysubgroup as ps
df = pd.read_csv("PsychophysioPreprocessed.csv", sep=',')
```

Next, we define the search space in which the algorithm should search. We remove the target column "Pleasantness_class", so that it is not considered as a feature and does not give us the rule "Pleasant implies pleasant". Setting the nbins parameter to 20 means that the algorithm must discretize the variable values into 20 classes. The greater this value, the greater the number of rules generated and the more precise the rules.

```
searchspace = ps.create_selectors(df, nbins=20,
ignore=['Pleasantness_class'])
```

We now need to specify the target class: in our case, it is the "Pleasantness_class" column, and we will start with the "unpleasant" odorants.

```
target = ps.BinaryTarget('Pleasantness_class', 'unpleasant')
```

Then, we create a Subgroup Discovery Task to identify the five best patterns (*result_set_size*) using the Weighted Relative Accuracy (*WRAcc*) value (*qf*). This task creates rules with a maximum of four conditions (*depth*) in the description of the subgroup. A condition can be written in different ways: either by a strict equality (e.g. "*Event*==0" for "the number of events is zero"), or by an interval (e.g. "Latency:[1.84:2.11[" for "the latency is between 1.84 and 2.11 ms"), or by a minimum or maximum value (e.g. "Rise-Time>=5.35" for "the rise time is greater than or equal to 5.35ms"). When several conditions are combined, they are separated by the "AND" operator, in this case, the different conditions must all be true for a trial to be included in the subgroup.

```
task = ps.SubgroupDiscoveryTask (df, target, searchspace,
result_set_size=5, depth=4, qf=ps.WRAccQF())
```

Now, we can extract the patterns. The algorithm, by default, is the "BEAM" search which performs a beam search exploration. It returns a "SubgroupDiscoveryResult" type object that can be converted into a dataframe and viewed in Table 3. We can then inspect the found subgroups and their characteristics.

```
result = ps.BeamSearch().execute(task)
unpleasant = result.to_dataframe()
```

We can also export the results to a csv file.

```
unpleasant.to_csv('unpleasant_result.csv', index=False, sep=',')
```

Then we can do the same search for "pleasant" odorants.

```
target = ps.BinaryTarget('Pleasantness_class', 'pleasant')
task = ps.SubgroupDiscoveryTask(df, target, searchspace,
    result_set_size=5, depth=4, qf=ps.WRAccQF())
result = ps.BeamSearch().execute(task)
pleasant = result.to_dataframe()
pleasant.to_csv('pleasant_result.csv', index=False, sep=',')
```

| Unpleasant pattern | | | | | | | |
|---|---|---|---|---|---|---|---|
| Pattern rank | Quality (WRAcc) | subgroup | size_sg | size_dataset | positives_sg | positives_dataset | lift |
| 1 | 0.006387 | Rise-Time>=5.35 | 114 | 2278 | 72 | 1130 | 1.25 |
| 2 | 0.003314 | Amplitude>=0.50 | 114 | 2278 | 65 | 1130 | 1.13 |
| 3 | 0.003282 | Event==2.0 | 259 | 2278 | 138 | 1130 | 1.06 |
| 4 | 0.002875 | Latency: [1.84:2.11[ | 114 | 2278 | 64 | 1130 | 1.11 |
| 5 | 0.002751 | Event ==0.0 AND Rise-Time>=5.35 | 59 | 2278 | 36 | 1130 | 1.21 |
| Pleasant pattern | | | | | | | |
| Pattern rank | Quality (WRAcc) | subgroup | size_sg | size_dataset | positives_sg | positives_dataset | lift |
| 1 | 0.005683 | Rise-Time: [1.92:2.26[ | 113 | 2278 | 69 | 1148 | 1.23 |
| 2 | 0.004285 | Event ==-1.0 | 319 | 2278 | 168 | 1148 | 1.06 |
| 3 | 0.003271 | Rise-Time: [3.03:3.41[ | 114 | 2278 | 64 | 1148 | 1.13 |
| 4 | 0.003053 | Latency>=3.20 | 115 | 2278 | 64 | 1148 | 1.12 |
| 5 | 0.002541 | Latency: [0.09:0.43[ | 73 | 2278 | 42 | 1148 | 1.16 |

**Table 3.** Top five results obtained with *pysubgroup* library (unpleasant and pleasant patterns are depicted separately).

## 2.3. Example of Results

The top five results obtained for the attribute pleasantness and its targets can be viewed in Table 3 in the same format as that generated by the pysubgroup library. The best model for unpleasant has the following rule: "Rise-Time> = 5.35 → unpleasant". The subgroup size is 114 (size_sg) which corresponds to 5% of the dataset (relative_size_sg or size_sg / size_dataset). Of these 114 trials, 72 are classified as "unpleasant" (positive_sg) and 42 as "pleasant" (size_sg-positive_sg). So we see that we do not get a 100% accurate rule (without error), but information like "if we have a Rise-Time> = 5.35, then the smell is more likely to be unpleasant". To know how correct this rule is, we use quality measures.

The quality of the model is calculated as the frequency of the rule in the subgroup relative to the frequency of the rule in the entire data set. The score generally used is the Weighted Relative Accuracy (*WRAcc*) value. The *WRAcc* value is the probability of having the target in the subgroup minus the probability of having the target in the data set, multiplied by the probability of the target in the data set.

$$WRAcc = size\_sg/size\_dataset \times (positives\_sg /size\_sg - positivesdataset /size\_dataset)$$

The *Wracc* value cannot exceed 0.25 or be less than -0.25. The higher the absolute value, the more exceptional the description feature in the subgroup compared to the rest of the data. A 0.25 value corresponds to the extreme case where the dataset is balanced and the subgroup – with a size of ½ size_dataset – uncovers all object from one target without uncovering the other target (i.e., true positive rate is 1 and false positive rate is 0). Another measure of quality is shown by the parameter *lift*. The *lift* is the probability of having the target (e.g. *pleasant* or *unpleasant*) in the subgroup divided by the probability of having the target in the whole dataset.

$$lift = (positive\_sg / size\_sg) / (positive\_dataset / size\_dataset)$$

If *lift* is less than 1, the rule is not interesting because the target is less frequent in the subgroup than outside. If we obtain a *lift* of 3, it means that the target is 3 times more frequent in the subgroup than in the whole data set.

We have data with a positive *WRAcc* and a *lift* greater than 1, which means that trends are found. The two quality measures are illustrated in Fig. 3. One way to be sure that a pattern is meaningful and not due to chance is to perform a bootstrap validation (see Note 4).

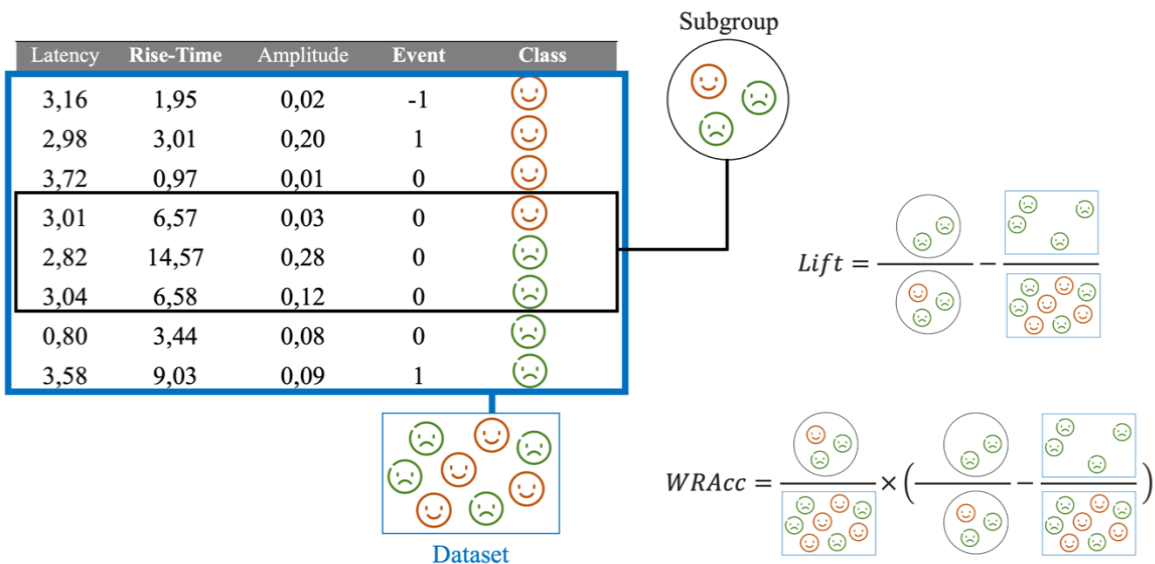

**Fig. 3**. Illustration of quality measures

2.4. Conclusions

Subgroup discovery methods have very little implementation in olfaction and food science, but when they have *(13, 19–21)*, they provided descriptive models linking physicochemical or physiological parameters to hedonic odor and aroma perceptions. This approach is promising and will be useful when seeking to describe perceptual phenomena and emotional reactions related to food perception from a physiological perspective.

## 4. Notes

*1. Machine learning analysis.* To analyze the link between physiological and emotional responses, it is possible to use other data science methods such as predictive approaches called supervised machine learning. Supervised machine learning can be used to achieve two goals. The first is to automate a process, for example, to learn directly about a person's perceived emotions without having to ask them. The second is to know how the classifier separates the different groups in order to understand the underlying neural or physiological process better. To do this, the algorithm must be intelligible or, if it is a black box, a method must be found to allow for explanation (22). The algorithm must also have a high prediction score, which can be difficult to achieve when trying to relate complex dimensions with high variability between individuals such as physiological responses and emotional responses. To make predictions, the scikit learn library available on Python is complete and offers multiple possibilities in terms of learning methods and algorithms. In R, there are libraries corresponding to the classifier to be used (e.g., e1071, rpart, klaR, kernlab, CORElearn, Rweka, tree, caret ramdomForest, nnet, glmnet, gbm, rath, ipred, ROCR, mboost). In KNIME (menu Analytics/Mining), you can find a series of learning algorithms including neural network, decision tree, logistic regression. In Orange, you can find classification algorithms in the Model menu (Logistic Regression, KNN, Random Forest, SVM...).

*2. Other data mining tools.* Besides Python, different tools and platforms exist to perform data science analysis: graphical interfaces such as Weka *(23)* and ELKI *(24)* or software such as KNIME *(25)* or Orange *(26)*, with which one can compose a specific workflow by assembling one after another modules performing a specific operation. There is also an easy-to-use software with a graphical interface called Cortana *(27)*. Cortana is also available as a plugin for KNIME. For those with good R skills, for subgroup discovery, one can use rsubgroup *(28)* on R.

*3. Discretization.* In the literature, ratings are usually discretized by dividing the scale as used into 2 or 3 or by dividing into percentiles of equal size. However, the way in which emotions and preferences are provided is unique to each person: some people rate using a wide range of nuances, whereas other do not. It is important to separate the ratings into several categories while limiting this subjectivity. Therefore, we propose a partitioning method that is neither equi-depth nor equi-width: use *KMeans* clustering on each subject independently. This algorithm allows partitioning into k clusters such that the distance between intra-cluster points is minimized and the inter-cluster distance is maximized allowing the subjective data to be partitioned into categories as different as possible. We do not recommend discretizing the scores by dividing the scale into equivalent spaces (e.g. 1-5 vs. 6-10, for 2 groups and a scale ranging from 1 to 10), or partitioning into median because this method does not always reflect a person's assessment strategy. Indeed, with these methods, odors perceived in a similar way can be found in different categories and very different odors in the same category *(13)*.

*4. Bootstrap validation.* To filter the patterns, we can add the *min_quality* parameter to the *SubgroupDiscoveryTask()* function in order to define the minimum accepted quality. However, this is not enough to guarantee that the pattern is significant. To validate the quality of a pattern, one can make a very large number of random draws of the same size as the pattern to be validated and look at the distribution of the quality measure of all the generated draws to verify that the pattern is outside the confidence interval of that distribution. For example, by calculating the *WRAcc* of 10,000 groups for each pattern discovered, with the same support as the current pattern, drawn at random (with discount between each draw). If its *WRAcc* is outside the 90% random distribution, then the pattern is validated and can be considered as interesting. This interval validation avoids flagging subgroups indicating a *WRAcc* likely to be observed by a random subset of entities. Two examples are shown in Fig. 4.

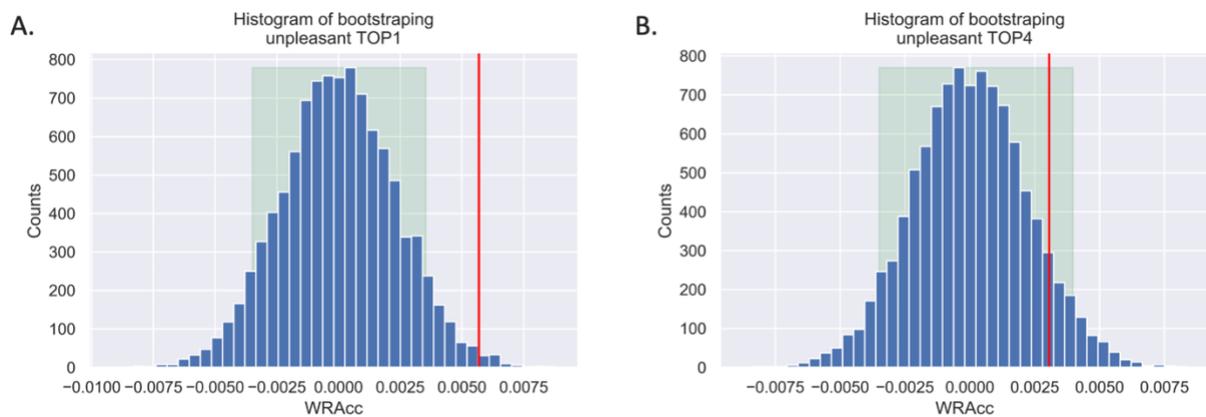

*Fig. 4. Examples of validation with on the left a validated pattern (A) and on the right a rejected pattern (B). The blue barplot corresponds to the distribution of the WRAcc of the random sample, the green rectangle to the confidence interval, and the red vertical line to the WRAcc of the pattern to be validated.*